\documentclass{webofc}
\usepackage[varg]{txfonts}   % Web of Conferences font
\newcommand{\ific}{Instituto de F\'{\i}sica Corpuscular (centro mixto CSIC-UV), 46980 Paterna, Valencia, Spain}
\newcommand{\ice}{Institute of Space Sciences (ICE, CSIC), Campus UAB, 08193 Barcelona, Spain}
\newcommand{\ieec}{Institut d'Estudis Espacials de Catalunya (IEEC), 08034 Barcelona, Spain}
\newcommand{\fias}{Frankfurt Institute for Advanced Studies, 60438 Frankfurt am Main, Germany}
\newcommand{\be}{\begin{equation}} \newcommand{\ee}{\end{equation}}
\newcommand{\ba}{\begin{array}{c}} \newcommand{\ea}{\end{array}}
\newcommand{\bea}{\begin{eqnarray}} \newcommand{\eea}{\end{eqnarray}}

\newcommand{\Tcc}{T_{cc}^+}
\newcommand{\bTcc}{T_{\bar c \bar c}^-}
\graphicspath{{images/}}

\begin{document}
\title{\boldmath Properties of the $T_{cc}(3875)^+$ and the $T_{\bar c \bar c}(3875)^-$ states in nuclear matter}

\author{\firstname{V\'ictor} \lastname{Montesinos}\inst{1}\fnsep\thanks{\email{vicmonte@ific.uv.es}} \and
        \firstname{Miguel} \lastname{Albaladejo}\inst{1} \and
        \firstname{Juan} \lastname{Nieves}\inst{1} \and 
        \firstname{Laura} \lastname{Tolos}\inst{2,3,4}
        % etc.
}

\institute{	\ific
\and
			\ice
\and
			\ieec
\and
			\fias
          }

\abstract{
Since its first detection, the interesting properties of the $T_{cc}(3875)^+$ have made it to be one of the most prominent tetraquark-like states up to date. In this work we present a joint analysis of the $\Tcc$ and its charge-conjugated partner in a dense nuclear medium. We start by considering both states as purely isoscalar $D^{\ast} D$ and $\overline{D}{}^{\ast} \overline{D}$ $S$-wave bound states, respectively. We use previous results for the in-medium $D$, $\overline{D}$, $D^\ast$ and $\overline{D}{}^\ast$ spectral functions to determine the modified two-meson amplitudes. We find important changes in the in-medium mass and width of both tetraquark-like resonances, which become more visible when increasing the nuclear density and molecular probabilities. The experimental confirmation of the found distinctive patterns will support the existence of molecular components in the $T_{cc}^+$ and $T_{\bar c\bar c}^-$ wave functions.
}
\maketitle
\section{Introduction}
\label{sec-intro}

In the past years the field of hadron spectroscopy has seen a new ``golden age'' with the discovery of many exotic states which do not fit the conventional quark models. This new revolution, which started with the $X(3872)$ \cite{Belle:2003nnu}, continues to see more new and interesting states to this day. One such state is the doubly-charmed $T_{cc}(3875)^+$ \cite{LHCb:2021vvq,LHCb:2021auc}, observed in the $D^0D^0 \pi^+$ mass distribution with a mass of $m_{\rm{thr}} + \delta m_{\rm{exp}}$, being $m_{\text{thr}} = 3875.09\,\text{MeV}$ the $D^{*+} D^0$ threshold and $\delta m_{\rm{exp}} = -360 \pm 40^{+4}_{-0}\,\text{keV}$, and a width $\Gamma = 48 \pm 2^{+0}_{-14}\,\text{keV}$ \cite{LHCb:2021auc}. Among the possible models, the molecular interpretation  is being supported due to its closeness to the $D^0D^{*+}$ and $D^+D^{*0}$ thresholds.

More experimental input on this state is thus very welcome in order to determine its internal structure. Recent works have studied the femtoscopic correlation functions of the $D^0D^{*+}$ and $D^+D^{*0}$ channels in heavy-ion collisions (HICs) \cite{Kamiya:2022thy,Vidana:2023olz}. Another possible way to gain some insight into the nature of the $\Tcc$ is to study its behavior under extreme temperature or density conditions, as has already been performed for the $X(3872)$ \cite{Albaladejo:2021cxj}. In this work we have followed the previous work of Ref.~\cite{Albaladejo:2021cxj} in order to study the behavior of the $\Tcc$ and its charge conjugated partner the $\bTcc$ in a dense nuclear environment, with the objective of analyzing the finite-density regime of HICs in experiments such as CBM at FAIR.

\section{Formalism}
\label{sec-formalism}

%In this section we present a brief summary of the formalism used to produce the results. More details can be found in Ref.~\cite{Montesinos:2023qbx}.

%In this section we briefly summarize the formalism of Ref.~\cite{Montesinos:2023qbx}.  Let's recall that the $\Tcc$ state has quantum numbers $I(J^P)=0(1^+)$ and is seen in the corresponding $D^\ast D$ channel.\footnote{The $\Tcc$  isospin quantum number still needs confirmation, though several analyses point to the isoscalar hypothesis as no signal is seen in the $D^{\ast +} D^+$ spectrum \cite{LHCb:2021vvq,LHCb:2021auc,Albaladejo:2021vln}.} Therefore, we start by considering $D^\ast D$ scattering in vacuum in the $I(J^P)=0(1^+)$ channel. We also take the exact isospin limit in which the masses of the $D^+$ and the $D^0$ mesons are identical. In this limit, both isospin channels decouple and we are left with a single-channel Bethe-Salpeter equation for the unitary $T$ matrix, which in the on-shell approximation reads
In this section we briefly summarize the formalism of Ref.~\cite{Montesinos:2023qbx} where this exotic state is generated as a result of $D^\ast D$ interaction.  The $\Tcc$ state is seen in the $D^0D^0\pi^+$ spectrum, right below the $D^{\ast+}D^0$ threshold, and has most probably quantum numbers $I(J^P)=0(1^+)$. Therefore, we start by considering $D^\ast D$ scattering in vacuum in this channel. We work in the exact isospin limit for the $D^{(\ast)}$ masses. In this limit, both isospin channels decouple and we are left with a single-channel Bethe-Salpeter equation, which in the on-shell approximation has as solution the unitary $T$ matrix
\be
\label{e-bethesalpeter}
T^{-1}(s)=V^{-1}(s)-\Sigma(s).
\ee
In the previous expression $V$ is a contact potential, and $\Sigma(s)$ is the $D^\ast D$ loop function. We consider two energy-dependent potential families:
\be\label{e-potentials}
V_A(s) = a + b s, \quad
V_B(s) = \frac{1}{c + d s},
\ee
with $a$, $b$, $c$ and $d$ parameters that can be fixed through Eq.~\eqref{e-bethesalpeter} by imposing that the mass of the $\Tcc$ in vacuum is $m_0$ and that the coupling of the $\Tcc$ to the $D^\ast D$ channel is $g_0$:
\be \label{e-parameterconditions}
T^{-1}(m_0^2) = 0 , \quad
\frac{dT^{-1}(s)}{ds}\Big |_{s=m^2_0} = \frac{1}{g^2_0} , \quad P_0 = -g_0^2 \Sigma^\prime(m_0^2) .
\ee
We also take into account Weinberg compositeness condition \cite{Weinberg:1965zz} (last equality in Eq.~\eqref{e-parameterconditions}), re-discussed in \cite{Gamermann:2009uq}, in order to analyze our results in terms of the molecular probability $P_0$ instead of the vacuum coupling $g_0$. Given that we are working in the exact isospin limit, the physical $\Tcc$ mass cannot be used, and instead we take $\delta m = -800$ keV with respect to the $D^*D$ threshold~\cite{Albaladejo:2021vln}. We do not fix $g_0$ or $P_0$ to any given value. This allows us to discuss our results in terms of these parameters.

At leading order, the effects of the nuclear medium on the scattering amplitude enter only through the meson loop function in Eq.~\eqref{e-bethesalpeter} (the potential is left unchanged):
\be\label{e-inmediumloopfunction}
\Sigma(s;\, \rho) = i \int \frac{d^4q}{(2\pi)^4} \Delta_{D^\ast}(P-q;\, \rho) \Delta_D(q;\, \rho) ,
\ee
where the in-medium mesonic propagator can be written as
\be
\Delta_U(q\,;\rho)  = \frac{1}{(q^0)^2-\omega_U^2(\vec{q}^{2}) - \Pi_U(q^0,\, \vec{q};\,\rho)}  = \int_0^\infty d\omega \left( \frac{S_{\!U}(\omega,\, \lvert \vec{q} \rvert;\,\rho)}{q^0 - \omega + i\varepsilon} - \frac{S_{\!\bar{U}}(\omega,\lvert \vec{q} \rvert;\,\rho)}{q^0 + \omega - i\varepsilon} \right), 
\ee
being $\Pi_U(q^0,\, \vec{q};\,\rho)$ the meson self energy and $S_{\!U}(\omega,\, \lvert \vec{q} \rvert)$ its spectral function. We use the model developed in Refs.~\cite{Tolos:2009nn,Garcia-Recio:2010fiq,Garcia-Recio:2011jcj} for computing the meson self energies. In order to regularize the loop integral of Eq.~\eqref{e-inmediumloopfunction} we use a sharp cutoff over three momentum of $\Lambda=700$ MeV. Further details on the computation of Eq.~\eqref{e-inmediumloopfunction} can be found in Refs.~\cite{Albaladejo:2021cxj,Montesinos:2023qbx}.

%The discussion for the $\bTcc$ runs parallel to all that has been said in this section regarding the $\Tcc$, but considering the meson pair $\overline{D}{}^{\ast} \overline{D}$ instead of $D^\ast D$. The fundamental difference lies in the anti-meson self energies, which are quite distinct from the meson ones owing to the contrasting $\overline{D}{}^{(\ast)}N$ and $D^{(\ast)}N$ interactions \cite{Tolos:2009nn,Garcia-Recio:2010fiq,Garcia-Recio:2011jcj}.

The discussion for the $\bTcc$ runs parallel to that for the $\Tcc$, but considering the meson pair $\overline{D}{}^{\ast} \overline{D}$ instead of the $D^\ast D$ one. The distinction arises from the anti-meson self energies, which divert significantly from those of mesons due to the quite different $\overline{D}{}^{(\ast)}N$ and $D^{(\ast)}N$ interactions \cite{Tolos:2009nn,Garcia-Recio:2010fiq,Garcia-Recio:2011jcj}.
%difference lies in the anti-meson self energies, quite different from the meson ones due to the contrasting $\overline{D}{}^{(\ast)}N$ and $D^{(\ast)}N$ interactions \cite{Tolos:2009nn,Garcia-Recio:2010fiq,Garcia-Recio:2011jcj}.

\section{Results}
\label{sec-results}

%In this section we present a comparison between the results obtained in $D^*D$ channel (where the $\Tcc$ pole is found) and the charge conjugated $\overline{D}{}^{\ast} \overline{D}$ channel (where the $\bTcc$ is found). We consider densities up to the normal nuclear density $\rho_0=0.17$ fm$^{-3}$. 

Here we compare the results obtained in the $D^*D$ and $\overline{D}{}^{\ast} \overline{D}$ channels, where the $\Tcc$ and $\bTcc$ poles are respectively found, for densities up to $\rho_0=0.17$ fm$^{-3}$ (normal nuclear density). 

\begin{figure*}[ht]
    \centering
    \includegraphics[width=.45\textwidth]{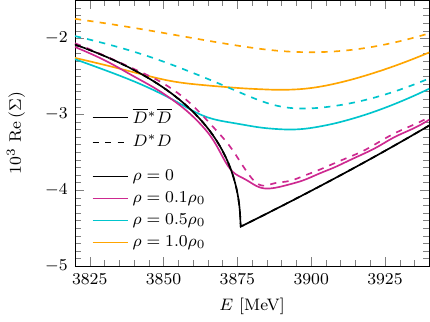}
    \hspace{5mm}
    \includegraphics[width=.45\textwidth]{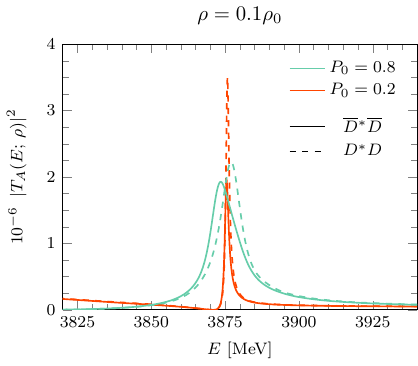}
    \includegraphics[width=.45\textwidth]{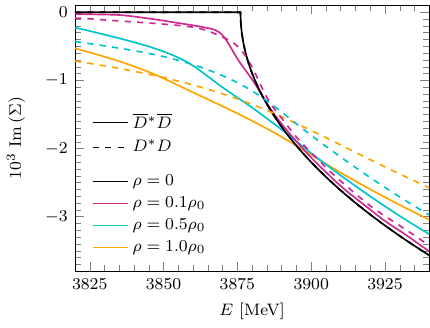}
    \hspace{5mm}
    \includegraphics[width=.45\textwidth]{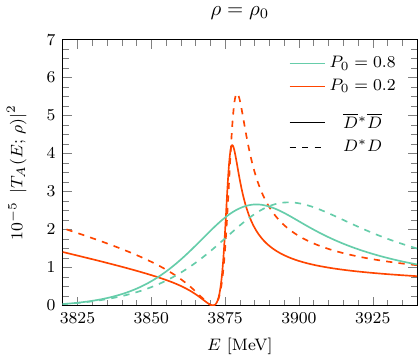}
    \caption{Left column: real (top) and imaginary (bottom) parts of the $\overline{D}{}^{\ast} \overline{D}$ (solid lines) and $D^*D$ (dashed lines) loop functions for different values of the nuclear medium density as a function of the center-of-mass energy $E$ of the meson pair. Right column: in-medium $\overline{D}{}^* \overline{D}$ (solid lines) and $D^* D$ (dashed lines) modulus square amplitudes obtained by solving Eq.~\eqref{e-bethesalpeter} using the $V_A(s)$ potential, for molecular probabilities $P_0=0.2$ (orange) and $P_0=0.8$ (blue), and for different nuclear densities.}
    \label{fig:Loop-TComparison}
\end{figure*}
In the left column of Fig.~\ref{fig:Loop-TComparison} we plot the real and imaginary parts of the $\overline{D}{}^{\ast} \overline{D}$ and $D^*D$ loop functions. The in-medium loop functions are smooth at threshold, and they develop an imaginary part below threshold. This accounts for the opening of new decay channels for the charmed mesons due to their interaction with the nucleons of the medium. Comparing both loop functions we see that they exactly coincide in vacuum, owing to charge conjugation symmetry, but when increasing density they stray apart. This different behavior is due to the very different $D^{(\ast)}N$ and $\overline{D}{}^{(\ast)}N$ interactions. Focusing now on the real part of the loop functions we find that it is more negative for $\overline{D}{}^{\ast} \overline{D}$ than for $D^*D$ scattering. This can be naively interpreted as the medium generating more repulsion in the $D^*D$ interaction than in the $\overline{D}{}^{\ast} \overline{D}$ one, but there is no clear answer as the imaginary part is not negligible \cite{Albaladejo:2021cxj,Montesinos:2023qbx}.

On the right column of Fig.~\ref{fig:Loop-TComparison} we plot the modulus squared of the in-medium $\overline{D}{}^\ast \overline{D}$ and $D^\ast D$ amplitudes for different densities and molecular probabilities, using the $V_A$ potential family presented in Eq.~\eqref{e-potentials}. Results using the $V_B$ potential are very similar to those stemming from $V_A$, specially for high molecular components~\cite{Albaladejo:2021cxj,Montesinos:2023qbx}. We find that the $\Tcc$ and $\bTcc$, which were bound states in vacuum, develop some width when embedded in the nuclear medium. Both states can be easily told apart in the high $P_0$ scenarios, and become more difficult to differentiate for small $\rho$ and $P_0$. Lastly, in the $P_0=0.8$ case we can clearly see that the $\Tcc$ peak sits to the right of the $\bTcc$ one, pointing to a more repulsive interaction in the $D^\ast D$ channel than in the $\bar{D}{}^\ast \bar{D}$ one, as we already mentioned when discussing the loop functions.

\section{Conclusions}
\label{sec-conclusions}

We have studied the behavior of the $T_{cc}(3875)^+$ and the $T_{\bar c \bar c}(3875)^-$ in a dense nuclear environment and have found that the medium effects are sizable when considering large values of the molecular probability, as opposed to the scenario where the molecular component is small. More specifically, we have observed an important increase in the width of the states and a noticeable shift in their masses. Furthermore, and due to the different $D^{(\ast)}N$ and $\overline{D}{}^{(\ast)}N$ interactions, we have seen that the $\Tcc$ and the $\bTcc$ states behave distinctly in the medium, the former shifting towards higher masses than the latter and becoming broader. If the $\Tcc$ and $\bTcc$ were to be predominantly compact objects, their density behavior, although distinct from one another, would likely be very different from the one found in the present work. All in all, we conclude that an interesting way to discern the molecular nature of the tetraquark-like $T_{cc}(3875)^+$ and $T_{\bar c \bar c}(3875)^-$ is to experimentally determine its density pattern in a dense nuclear environment. This could be tested at experiments such as PANDA (FAIR) with fixed nuclear targets like $\bar p$-nuclei, or in HICs at the CBM (FAIR).

\section*{Acknowledgements}
This work was supported by the Spanish Ministerio de Ciencia e Innovaci\'on (MICINN) under contracts No.\, PID2019-110165GB-I00 and No.\, PID2020-112777GB-I00, by Generalitat Valenciana under contract PROMETEO/2020/023, and from the project CEX2020-001058-M Unidad de Excelencia ``Mar\'{\i}a de Maeztu"). This project has received funding from the European Union Horizon 2020 research and innovation programme under the program H2020-INFRAIA-2018-1, grant agreement No.\,824093 of the STRONG-2020 project.  M.\,A. and V.\,M.~are supported through Generalitat Valenciana (GVA) Grants No.\,CIDEGENT/2020/002 and ACIF/2021/290, respectively. %and thanks the warm support of ACVJLI. 
L.\,T. also acknowledges support from the CRC-TR 211 `Strong-interaction matter under extreme conditions'- project Nr. 315477589 - TRR 211 and from the Generalitat de Catalunya under contract 2021 SGR 171.

%
% BibTeX or Biber users please use (the style is already called in the class, ensure that the "woc.bst" style is in your local directory)
\bibliography{references}

\begin{thebibliography}{13}

\bibitem{Belle:2003nnu}
S.K. Choi et~al. (Belle), Phys. Rev. Lett. \textbf{91}, 262001 (2003),
  \texttt{hep-ex/0309032}

\bibitem{LHCb:2021vvq}
R.~Aaij et~al. (LHCb), Nature Phys. \textbf{18}, 751 (2022),
  \texttt{2109.01038}

\bibitem{LHCb:2021auc}
R.~Aaij et~al. (LHCb), Nature Commun. \textbf{13}, 3351 (2022),
  \texttt{2109.01056}

\bibitem{Kamiya:2022thy}
Y.~Kamiya, T.~Hyodo, A.~Ohnishi, Eur. Phys. J. A \textbf{58}, 131 (2022),
  \texttt{2203.13814}

\bibitem{Vidana:2023olz}
I.~Vidana, A.~Feijoo, M.~Albaladejo, J.~Nieves, E.~Oset (2023),
  \texttt{2303.06079}

\bibitem{Albaladejo:2021cxj}
M.~Albaladejo, J.M. Nieves, L.~Tolos, Phys. Rev. C \textbf{104}, 035203 (2021),
  \texttt{2102.08589}

\bibitem{Montesinos:2023qbx}
V.~Montesinos, M.~Albaladejo, J.~Nieves, L.~Tolos, Phys. Rev. C \textbf{108},
  035205 (2023), \texttt{2306.17673}

\bibitem{Weinberg:1965zz}
S.~Weinberg, Phys. Rev. \textbf{137}, B672 (1965)

\bibitem{Gamermann:2009uq}
D.~Gamermann, J.~Nieves, E.~Oset, E.~Ruiz~Arriola, Phys. Rev. D \textbf{81},
  014029 (2010), \texttt{0911.4407}

\bibitem{Albaladejo:2021vln}
M.~Albaladejo, Phys. Lett. B \textbf{829}, 137052 (2022), \texttt{2110.02944}

\bibitem{Tolos:2009nn}
L.~Tolos, C.~Garcia-Recio, J.~Nieves, Phys. Rev. C \textbf{80}, 065202 (2009),
  \texttt{0905.4859}

\bibitem{Garcia-Recio:2010fiq}
C.~Garcia-Recio, J.~Nieves, L.~Tolos, Phys. Lett. B \textbf{690}, 369 (2010),
  \texttt{1004.2634}

\bibitem{Garcia-Recio:2011jcj}
C.~Garcia-Recio, J.~Nieves, L.L. Salcedo, L.~Tolos, Phys. Rev. C \textbf{85},
  025203 (2012), \texttt{1111.6535}

\end{thebibliography}
%\begin{thebibliography}{}
%Journal Author, Journal \textbf{Volume}, page numbers (year)
%Book Author, \textit{Book title} (Publisher, place, year) page numbers
% etc
%\end{thebibliography}

\end{document}

% end of file template.tex

<div id='footer'><table width='100%'><tr><td class='right'><a href='http://fusioninventory.org/'><span class='copyright'>FusionInventory 9.1+1.0 | copyleft <img src='/glpi/plugins/fusioninventory/pics/copyleft.png'/>  2010-2016 by FusionInventory Team</span></a></td></tr></table></div>